\documentclass[structabstract]{aa} 
\usepackage{graphicx}
\usepackage{txfonts}

\newcommand{\Mjup}{M$_\mathrm{J}$}
\newcommand{\Rjup}{R$_\mathrm{J}$}

\newcommand{\ratioone}{(R_p/R_\star)_{3.6\,\mu\mathrm{m}}}

\newcommand{\etal}{et al.}
\newcommand{\sn}{{\rm S/N}}
\newcommand{\spitzer}{\emph{Spitzer}}
\newcommand{\micron}{$\mu$m}
\newcommand{\hd}{HD~189733b}
\newcommand{\HOO}{H$_2$O}
\newcommand{\CHQ}{CH$_4$}
\newcommand{\COD}{CO$_2$}
\begin{document}

\title{Transit spectrophotometry of the exoplanet HD189733b. \\II. New \emph{Spitzer} observations at 3.6~\micron}

 \titlerunning{New \emph{Spitzer} observations at 3.6~\micron} %

\author{J.-M. D\'esert\inst{1,2} \and D.
   Sing\inst{1,3} \and A. Vidal-Madjar\inst{1} \and G. H\'ebrard\inst{1} \and D. Ehrenreich\inst{4} \and A. Lecavelier des Etangs\inst{1} \and V. Parmentier\inst{1} \and R. Ferlet\inst{1}  \and G. W. Henry\inst{5}}

   \offprints{desert@iap.fr}

\institute{
  Institut d'Astrophysique de Paris, UMR7095 CNRS, Universit\'e Pierre \& Marie Curie, 98bis boulevard Arago, 75014 Paris, France
\email{jdesert@cfa.harvard.edu}
 \and
       Harvard-Smithsonian Center for Astrophysics, Cambridge, MA 02138
         \and
 Astrophysics Group, School of Physics, University of Exeter, Stocker Road, Exeter EX4 4QL, UK
         \and         
       Laboratoire d'Astrophysique de l'Observatoire de Grenoble, Université Joseph Fourier, CNRS (UMR 5571), BP 53, 38041 Grenoble cedex 9, France
 \and
          Center of Excellence in Information Systems, Tennessee State University,
          3500 John A. Merritt Blvd., Box 9501, Nashville, TN 37209, USA
}

   \date{Received ??, 2009; accepted ????, 2010}

 \abstract
   {We present a new primary transit observation of the hot-jupiter
HD\,189733b, obtained at 3.6~\micron\ with the Infrared Array
Camera (IRAC) onboard the \emph{Spitzer Space Telescope}. Previous
measurements at 3.6 microns suffered from strong systematics and
conclusions could hardly be obtained with confidence on the water
detection by comparison of the 3.6 and 5.8 microns observations.}
  {We aim at constraining the atmospheric structure and composition of the planet and improving over previously derived parameters.}
 {We use a high-$S/N$ \emph{Spitzer} photometric transit light curve to improve the precision of the near infrared radius of the planet at 3.6~\micron. The observation has been performed using high-cadence time series
integrated in the subarray mode. We are able to derive accurate
system parameters, including planet-to-star radius ratio, impact
parameter, scale of the system, and central time of the transit
from the fits of the transit light curve. We compare the results
with transmission spectroscopic models and with results from
previous observations at the same wavelength.}
   {We obtained the following system parameters of $R_p/R_\star=0.15566^{+0.00011}_{-0.00024}$, $b=0.661^{+0.0053}_{-0.0050}$, and
$a/R_\star=8.925^{+0.0490}_{-0.0523}$ at 3.6~\micron. These measurements are three times more accurate than previous studies at this wavelength because they benefit from greater observational efficiency and less statistic and systematic errors. Nonetheless, we find that the radius ratio has to be corrected for stellar activity and present a method to do so using ground-based long-duration photometric follow-up in the V-band. The resulting planet-to-star radius ratio corrected for the stellar variability is in agreement with the previous measurement obtained in the same bandpass (D\'esert et al. 2009). We also discuss that water vapour could not be evidenced by comparison of the planetary radius measured at 3.6 and 5.8~\micron, because the radius measured at 3.6~\micron\ is affected by absorption by other species, possibly Rayleigh scattering by haze.}
  {}


   \maketitle

%

\section{Introduction}

HD\,189733b is a transiting hot jupiter that orbits a small,
bright, main-sequence K2V star ($K = 5.5$) and produces deep
transits of $\sim$2.5\% (Bouchy \etal\ 2005). The planet has a
mass of $M_p = 1.13$~Jupiter mass (\Mjup), a radius of $R_p =
1.16$~Jupiter radius (\Rjup) in the visible (Bakos \etal\ 2006b;
Winn \etal\ 2007), and a brightness temperature ranging between
950 and 1220~K (Knutson et al. 2007). This implies a large
atmospheric scale height (200~km), allowing transits to
characterize the chemical composition and vertical structure of
the atmosphere.

Transmission spectroscopy during transit allows the upper part of
the atmosphere to be probed, down to altitudes where it becomes
optically thick. Therefore, one can infer the composition of the
atmosphere by measuring the wavelength-dependent planetary radius
(Seager \& Sasselov 2000; Burrows et al. 2001; Hubbard et al.
2001). Pioneering observational work using this method on
HD\,209458b with the \emph{Hubble Space Telescope (HST)} has led
to the first detection of an exoplanetary atmosphere (Charbonneau
\etal\ 2002) and its escaping exosphere (Vidal-Madjar \etal\ 2003;
2004; 2008; Ballester et al. 2007). The optical transmission
spectra of this planet shows evidence for several different layers
of Na (Sing et al. 2008a,b) as well as Rayleigh scattering by
molecular hydrogen (Lecavelier des Etangs et al. 2008b) and the
possible presence of TiO/VO (D\'esert et al. 2008). Sodium has
been detected in the atmosphere of HD\,189733b, with ground-based
observations (Redfield et al. 2008). Using the \emph{Advanced
Camera for Survey (ACS)} aboard \emph{HST}, Pont et al. (2007, 2008)
detected atmospheric haze high altitude, interpreted as Rayleigh
scattering, possibly by small particles (Lecavelier des Etangs et
al. 2008a). The detection of water and methane reported by Swain
et al. (2008) in spectroscopic observations between 1.5 and
2.5~\micron\ has been challenged by Sing et al. (2009) with new
\emph{HST} observations at 1.66 and 1.8~\micron. More recently, Lecavelier des Etangs et al. (2010) reported the presence of an extended exosphere of atomic hydrogen surrounding the planet using far-UV observations.

The \spitzer\ \emph{Space Telescope} (Werner et al. 2004) has
demonstrated its potential to probe exoplanetary atmospheres
through emission and transmission spectra. Multi-wavelength
observations of the eclipses (secondary transits) of HD\,189733b
behind the star have allowed the study of the atmosphere's
emission spectrum (Charbonneau et al. 2005; Deming et al. 2005, 2006, 2007). This planet has been observed extensively during secondary transits and throughout the phases of its orbit
(Knutson et al. 2007, 2009). Strong water (\HOO) absorption in the
dayside emission spectrum has been detected along with possible
variability (Grillmair et al. 2007, 2008). The likely presence of
carbon monoxide (CO; Charbonneau et al. 2008), and carbon dioxide
(\COD; Swain et al. 2009) have also been reported.

Knutson et al. (2007) obtained the first accurate near infrared
(NIR) transit measurements for this planet at 8.0~\micron\ using
the Infrared Array Camera (IRAC; Fazio et al. 2004) aboard
\spitzer. The search for molecular spectroscopic signatures by
comparing two photometric bands with \spitzer\ has been attempted
by Ehrenreich et al. (2007). This study concluded that
uncertainties in the measurements were too large to draw firm
conclusions on the detection of water at high altitudes. Yet,
using the same data set but a different analysis (Beaulieu et al. 2008), Tinetti et al. (2007) claimed to detect
the presence of atmospheric water vapour from comparison of the
absorption at 3.6~\micron\ and 5.8~\micron.

We recently performed a consistent and complete study of these
observations obtained with the four IRAC channels (D\'esert et al.
2009), including a detailed assessment of systematics. We
concluded that there is no excess absorption at 5.8~\micron\
compared to 3.6~\micron. We also derived a slightly larger radius
at 4.5~\micron\ that cannot be explained by \HOO\ or Rayleigh
scattering. We interpreted this small absorption excess as due to
the presence of CO molecules (D\'esert et al. 2009). This is
consistent with the low level of emission measured from the
planetary eclipse at the same wavelength (Charbonneau et al.
2008).

Here we present new Spitzer/IRAC observations of a primary transit
of HD\,189733b at 3.6~\micron. This work is part of our ongoing
efforts to characterize the transmission spectrum of HD189733
using space-based observatories (D\'esert et al. 2009; Sing et al.
2009). 
The present work focuses on new \emph{Spitzer/IRAC} observations of
the planetary transit, where the photons were accumulated in
subarray mode. The data presented here consists of a high-cadence
time series, which are described in Sect.~\ref{sec:obsred} and
analyzed in Sect.~\ref{sec:ana}. We present a method to correct for the stellar variability in Sect.~\ref{sec:res}. The photometric precision achieved with this observation allows us to derive accurate measurements of the planetary radius and orbital parameters. Our results together with a comparison with previous studies and theoretical
predictions are also given in Sect.~\ref{sec:res}.

\section{Observations and data reduction}
\label{sec:obsred}

\subsection{Observations}
\label{sec:obs}

We obtained \spitzer\ General Observer's time in Cycle 3 and 4 (PI:
A. Vidal-Madjar; program IDs 30590 \& 40732), in the four
\emph{Spitzer/IRAC} bandpasses. In total, three primary transits
of \hd\ were observed. Our primary scientific objectives were to
detect the main gaseous constituents (\HOO\ and CO) of the
atmosphere of this hot-Jupiter.

Our first observations of the HD~189733 system (hereafter visit 1)
were acquired simultaneously at 3.6 and 5.8~\micron\ (channels 1
and 3) on 2006 October 31. Visit~2 was completed 2007 November 23
with measurements at 4.5 and 8~\micron\ (channels 2 and 4). For
both visits, the system was observed in IRAC's stellar mode for
4.5 hours, of which 1.8h was during planetary transit. The
observations were binned into consecutive subexposures with a
cadence of 0.4 sec for channels 1 and 2 and 2.0 sec for channels 3
and 4. We obtained a total of 1936 frames for channels 1 and 3 and
1920 frames for channels 2 and 4. Details of the analysis of these
data (Visit~1 and 2) can be found in D\'esert et al. (2009).

Here we focus on the data obtained during Visit~3 on 2007 November 25 (cycle~4, program 40732) at 3.6~\micron\ (channel~1);
its analysis is presented in this paper. We used IRAC's
$32~\times~32$ pixel subarray mode (1.2'' $\times$ 1.2'' pixels)
for visit~3, which lasted 4.5 hours with 1.8h during transit. The
primary goal of this last visit was to obtain an accurate
measurement of the planetary radius at 3.6~\micron. The subarray mode has a greater efficiency than the stellar mode in collecting photon.

HD~189733 is brighter in channels 1 and 2 than in channels 3 and
4. For visit~1, the flux was 1,700~mJy at 3.6~\micron\, which is
close to the saturation limit with 0.4~s exposure time as used in
visit~1. IRAC's subarray mode can solve this issue by observing at
high speed cadence of 0.1~s with sets of 64 subframes taken
back-to-back with no gaps between subframes. During visit~3, we
obtained 1920 consecutive exposures, yielding to 122880 frames
with an effective exposure time of 0.08 sec and a frame interval
of 0.1 s. A delay of 8.3 sec occurs between consecutive exposures.
While planning the observation in the Astronomical Observing
Request (AOR) format, we carefully avoided placing the star near
dead pixels. We also did not dither the pointing so the target
remained on a given pixel throughout the observation. This
minimizes errors from imperfect flat-field corrections and, thus,
increases the relative photometric accuracy. This strategy has been used
successfully for several previous \emph{Spitzer} observing runs on
HD~189733 (Knutson et al. 2007; Ehrenreich et al. 2007; Agol et
al. 2008; D\'esert et al. 2009), on HD~149026 (Nutzman et al.
2009), on GJ 436 (Deming et al. 2007, Gillon et al. 2007, Demory et al. 2007), and more recently with Warm-Spitzer on HD~80606 (H\'ebrard et al. 2010).


\subsection{Data reduction}
\label{sec:red}

For visit~3, we used Basic Calibrated Data (BCD) frames, produced
by the standard IRAC calibration pipeline, to correct for dark
current, flat-fielding, and detector nonlinearity and to convert
the observations into flux units.  We are able to measure the
centroid position of a stellar image (computed by Gaussian
fitting) to a precision of $0.01$ pixel, using the DAOPHOT-type
Photometry Procedures, \texttt{GCNTRD}, from the IDL Astronomy
Library \footnote{{\tt http://idlastro.gsfc.nasa.gov/homepage.html}}. We find that the
centroid position varies by less than $10\%$ of a pixel during a
complete observation. We used the \texttt{APER} routine to
perform aperture photometry with a radius of $5$ pixels to
minimize the contribution of HD~189733B, which stands at an
angular separation of $11\farcs2$ (Bakos et al. 2006a).

The background level for each frame was measured with
\texttt{APER} as the median value of the pixels inside an annulus
centered on the star with inner and outer radii of $16$ and $18$
pixels, respectively. Typical background values are $10\pm0.2$
electrons per pixel remained compared to $\sim 110,000$ electrons
of total stellar signal. Therefore, photometric errors are not
dominated by fluctuations in the background. Short exposures yield
a typical signal-to-noise ratio ($S/N$) of $\sim 300$ per
individual observation.

After producing the photometric time series, we iteratively
selected and trimmed outliers greater than $4~\sigma$ by comparing
the $122,880$ photometric measurements to the best-fit transit
light curve model. This removed any remaining measurements
affected by transient hot pixels. In this way, we discarded $138$
frames, or approximately $0.1\%$ of the observations.  We also
discarded the first $2542$ measurements (planetary phase before
$-0.0472$), which are affected by a small ramp effect. Thus, our
cleaned light curve consists of 120,200 photometric measurements.
We then binned the transit light curve by a factor of four to
increase computing efficiency without losing information on the
pixel-phase effect (see below).  Our final cleaned, binned transit
light curve contains 30,050 data points.


\section{Analysis}
\label{sec:ana}

\subsection{Fitting the Transit Light Curve}
\label{sec:fit}

\begin{figure*}
\centering
\includegraphics[angle=90,width=18cm]{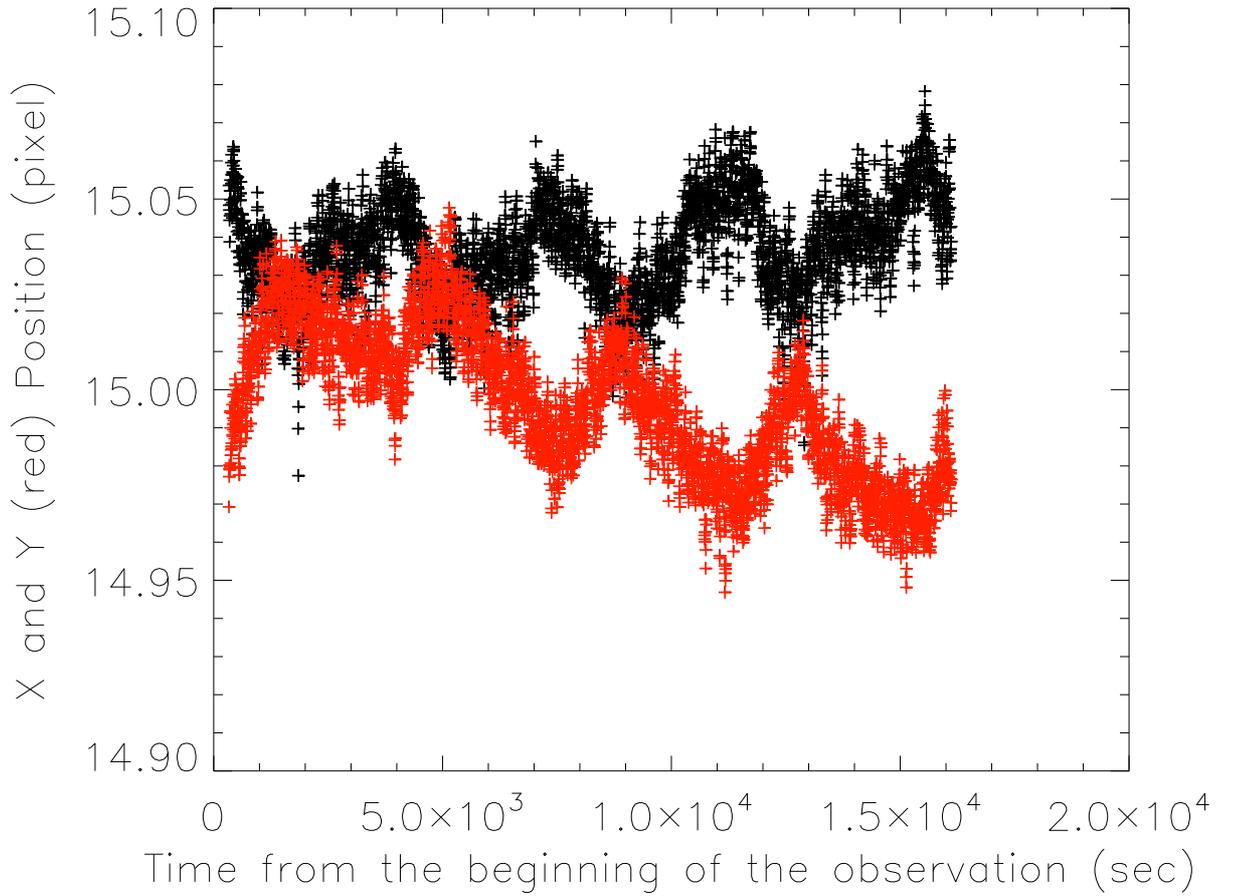}
\newline
\caption{X (black) and Y (red) position of the star centroid in
pixels as function of time (binned by 4). These positions are used
in the decorrelation function that is used the fit of the transit
light curve.} \label{fig:xypos}
\end{figure*}

\begin{figure*}
\centering
\includegraphics[angle=90,width=18cm]{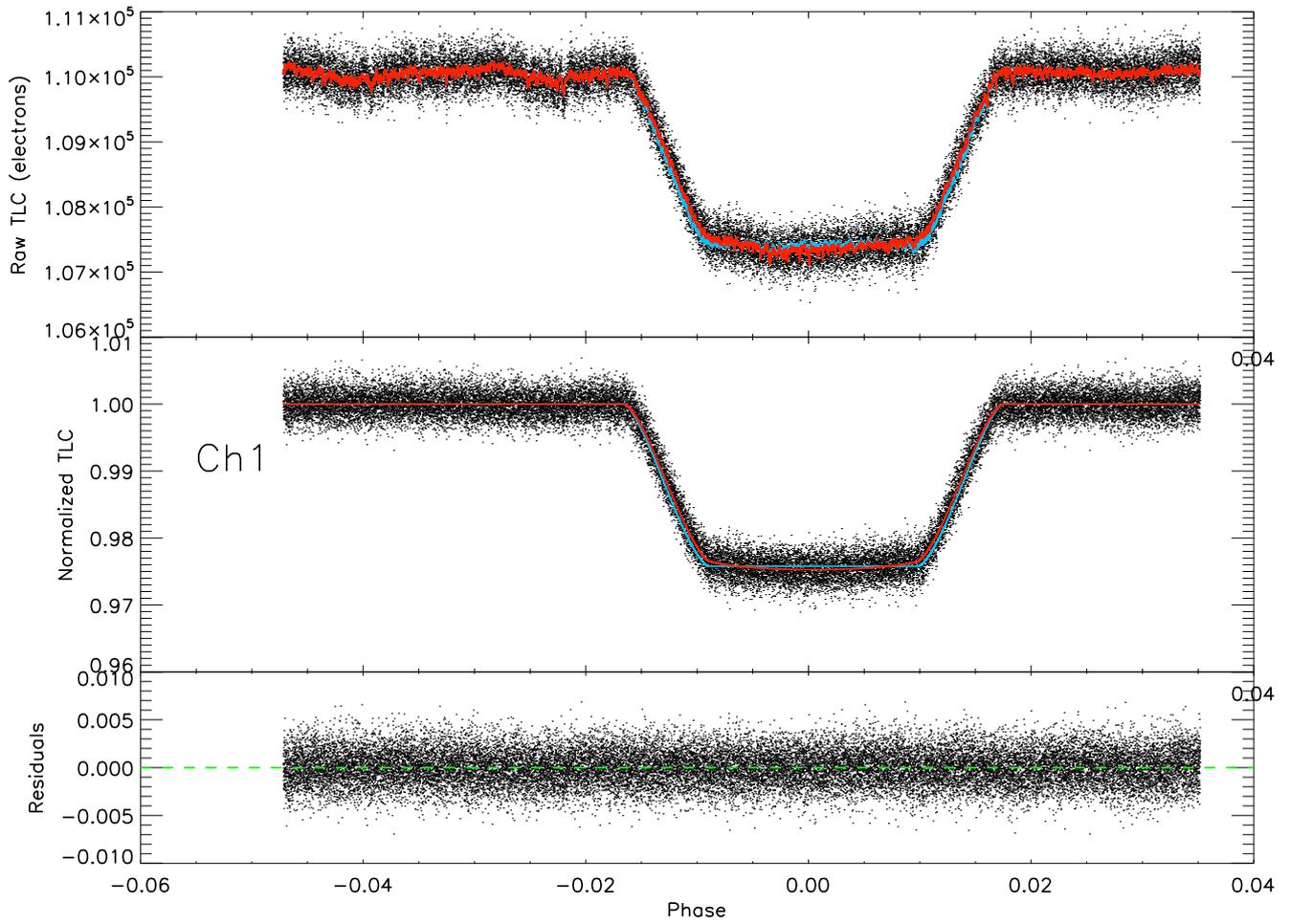}
\newline
\caption{Top panel: raw transit light curve obtained in subarray
mode binned by four with its fit overplotted in red. Pixel-phase
effect is clearly visible. Middle panel: Normalized transit light
curve. Bottom panel: residuals obtained from the fits of the
transit light curve.} \label{fig:tlc}
\end{figure*}

Due to instrumental effects, our measured out-of-transit flux
values are not constant.  Telescope pointing jitter results in fluctuations
of the stellar centroid position (Fig.~\ref{fig:xypos}), which, in
combination with intra-pixel sensitivity variations, produces
systematic noise in the raw light curves (Fig.~\ref{fig:tlc},
upper panel).  A description of this effect, known as the
pixel-phase effect, is given in the \spitzer/IRAC data handbook
(Reach et al.\ 2006, p.~50; see also Charbonneau et al. 2005). To
correct the light curve, we define a baseline function that is the
sum of a linear function of time and a quadratic function of the X
and Y centroid positions.  This function, with five parameters,
$K_i$, is described in D\'esert et al. (2009).

We modelled the transit light curve with 4 parameters: the
planet-star radius ratio, $R_p / R_\star$, the orbital semi-major
axis to stellar radius ratio (system scale), $a / R_\star$, the
impact parameter, $b$, and the time of mid transit, $T_c$.  We
used the IDL transit routine \texttt{OCCULTNL}, developed by
Mandel \& Agol (2002), to model the light curve. Our
limb-darkening corrections consist of three non-linear,
limb-darkening coefficients, as defined in Sect.~\ref{sec:ld} and
presented in Table~1. We also used a linear function of time as
the baseline (A$_j$, $j$=1,2) as presented in D\'esert et al.
(2009).

We used the \texttt{MPFIT} package\footnote{{\tt
http://cow.physics.wisc.edu/$\sim$craigm/idl/idl.html}} to perform
a Levenberg-Marquardt least-squares fit of the transit model to
our cleaned and binned observations.  The best-fit model was
computed over the whole parameter space ($R_p / R_\star$, $a /
R_\star$, $b$, $T_c$, $A_j$, $K_i$).  The baseline function
described above is combined with the transit light curve function
so the fit is constrained by $11$ parameters (4 for the transit
model, 2 for the linear baseline, and 5 for the pixel phase
effect); this gives $30,039$ degrees of freedom.

\subsection{Mean values and errors determination}
\label{sec:values}

We used the Prayer Bead method (Moutou et al. 2004, Gillon et al.
2007) to determine the mean value as well as the statistical and
systematical errors for the measured parameters.

As shown by Pont et al. (2006), the existence of low-frequency
correlated noise (dubbed red noise) between different exposures
must be considered to obtain a realistic estimation of the
uncertainties. To obtain an estimate of the systematic errors in
our observations we followed the method described by Gillon et al.
(2006) and derived the covariance from the residuals of the light
curve. In the Prayer Bead method, the residuals of the initial fit
are shifted systematically and sequentially by one frame, and then
added to the transit light curve model before fitting again. The
error on each photometric point is the same and is set to the
$rms$ of the residuals of the first fit obtained.

\begin{figure*}
\centering
\includegraphics[angle=90,width=18cm]{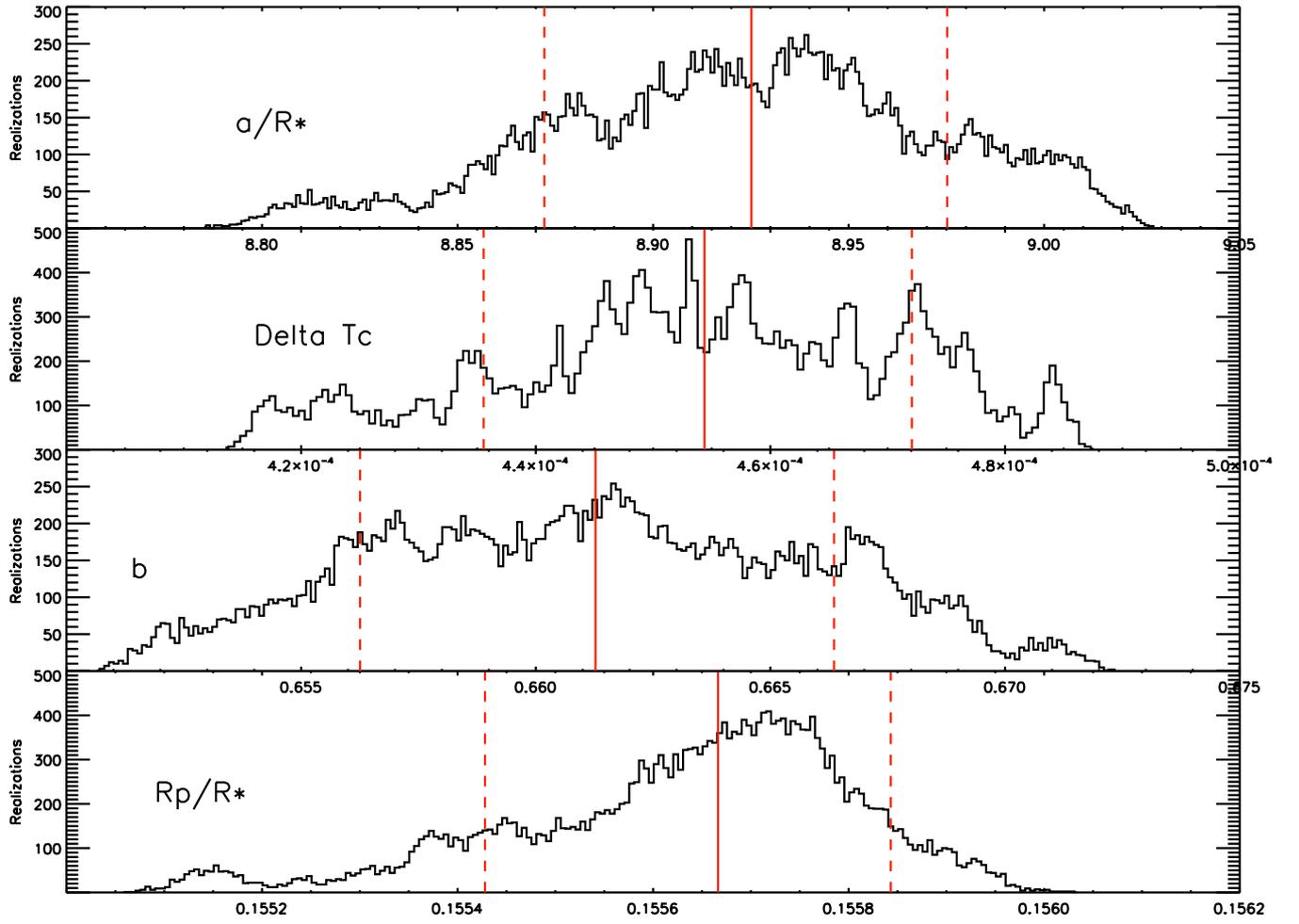}
\caption{Distributions obtained from the Prayer Bead method for each fitted parameters. Vertical continuous lines correspond to the median of the distribution. Vertical dashed lines
correspond to plus or minus 68\% of the distribution.}
\label{fig:prayerbead}
\end{figure*}

\begin{figure*}
\centering
\includegraphics[angle=90,width=18cm]{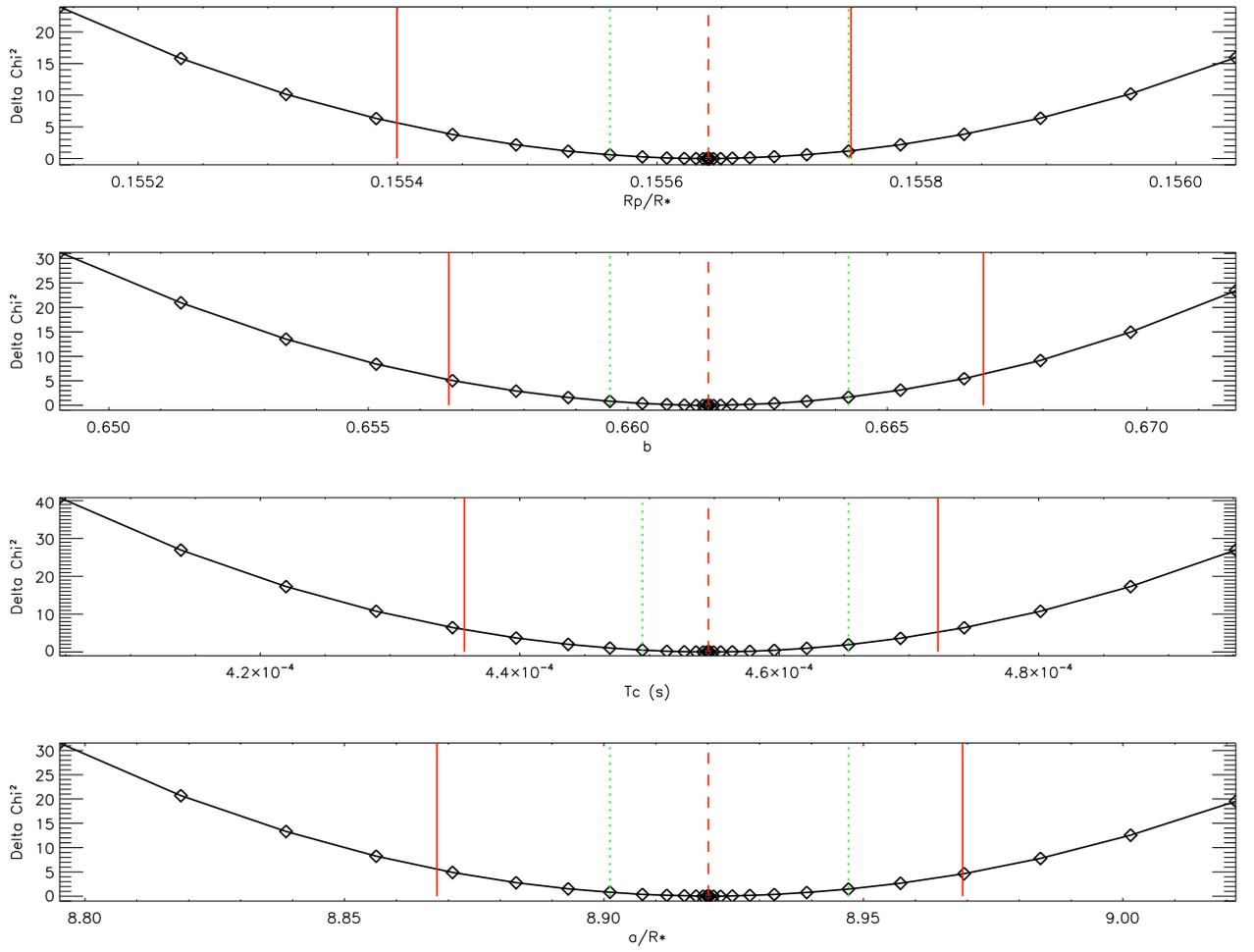}
\caption{Delta-Khi-square distribution of parameters. Vertical dashed
red lines correspond to the values for the minimal Khi-square and
vertical continuous red lines correspond to plus or minus
$1\sigma$ extracted with the prayer bead method. Vertical dotted
green lines correspond to plus or minus $1\sigma$ extracted with
the Khi-square method.} \label{fig:chisquare}
\end{figure*}

Totally, $30050$ shifts and fits of transit light curves were
produced to derive a set of parameters and to extract their
medians and their corresponding errors. We set our uncertainties
equal to the range of values containing $68\%$ of the points in
the distribution in a symmetric range about the median for a given
parameter (Fig.~\ref{fig:prayerbead}). 
Error bars estimated using the Prayer Bead method are generally larger than the ones obtained using other methods like ''delta khi square" (Fig.~\ref{fig:chisquare}) or Markov Chain Monte Carlo (MCMC) because they include the effect of the residual red noise. Therefore these error bars are likely the most conservative.


\section{Results and discussion}
\label{sec:res}

\begin{figure*}
\centering
\includegraphics[angle=90,width=18cm]{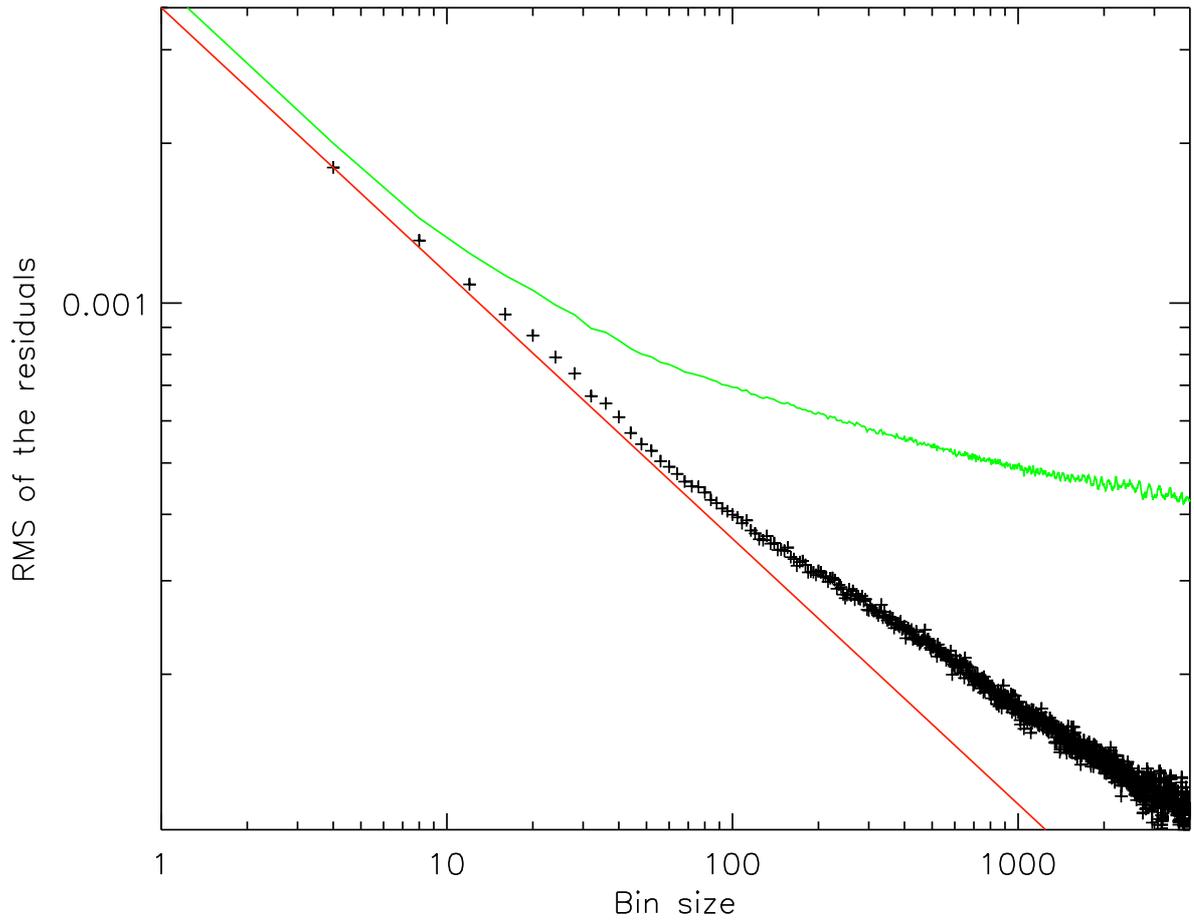}
\newline
\caption{Root-mean-square of binned residuals versus bin size. The
solid continuous red line is proportional to $N^{-1/2}$ and is
normalized to match the value for bin size $N = 4$. Top continuous
green curve correspond to bin residuals without pixel-phase
decorrelation. Black cross correspond to bin residuals with
pixel-phase effect decorrelation. Remaining red-noise become
apparent for bin larger than 50 data points (6.25 seconds).} \label{fig:binres}
\end{figure*}

The raw aperture photometry of the transit light curve
obtained in subarray mode binned by four is plotted in
Figure~\ref{fig:tlc}, together with the lightcurve decorrelated
from the dependence on instrumental parameters. We derived the
system parameters and their error bars in a consistent way using
the distribution obtained from the prayer bead method and present
the results in Table~2. From our fits, we evaluated the radius
ratio $R_p/R_*$, the impact parameter $b$, the system scale
$a/R_\star$ and the central time of the transit $T_c$ (Table~2).
We obtain a mean $\chi^2$ of $30024$ for $n=30039$ degrees of
freedom. However, we found that the parameters distributions are
far from being Gaussian (Fig.~\ref{fig:prayerbead}). This is
explained by the presence of a non negligible red-noise left after
corrections as revealed by comparing the $rms$ of the unbinned and
binned residuals (Fig.~\ref{fig:binres}) following the method
described by Gillon et al. (2006). This strenghtens the suggestion that error bars would have been smaller if estimated with methods other than the Prayer Bead.

\subsection{Signal-to-noise ratios}

The standard deviation of the difference between the best fit
model and the observed transit light curve is found to be $1.8
\times 10^{-3}$ (rms) on individual data points. This value
corresponds to a signal-to-noise of $560:1$ for a frame binned by
four; as expected, this is nearly twice the theoretical (photon
noise) signal-to-noise ratio for individual frames. We achieve an average of 91\% of the theoretical signal-to-noise on individual frames.

We also estimated the red-noise by comparing the $rms$ of residuals
with different bin sizes (Fig.~\ref{fig:binres}) and we find that
correlated systematics left after decorrelation of the pixel-phase
effect have an amplitude of $1.4 \times 10^{-4}$ in a bin of a $1000$ points. The amplitude of
the white ($\sigma_w$) and red ($\sigma_r$) noise are obtained
by solving the following system of equations:
\begin{eqnarray}
\sigma_4^2 &=&\frac{\sigma_w^2}{4} + \sigma_r^2\; \\
\sigma_{1000}^2 &=&\frac{\sigma_w^2}{1000} + \sigma_r^2\; \; ,
\end{eqnarray}
where $\sigma_4$ and $\sigma_{1000}$ are the standard deviation
taken over a sliding average of four and of one thousand points
respectively. We estimated the white noise amplitude, $\sigma_w =
0.003589$, and the low-frequency red noise amplitude, $\sigma_r =
0.0001397$.

\subsection{Transit timing}

The measured central time of the transit can be compared to the
expected transit time from a known ephemeris. We measure timing
residuals for the transit according to the ephemeris of Knutson et
al. (2009) ($P=2.21857578  \pm 0.00000080$ days, $Tc=2454399.23990
\pm 0.00017$ HJD). We find that the center of our transit is
$T_c=2454430.310594$ (HJD), which corresponds to an observed minus
calculated ($O-C$) transit time of $47 s \pm 4 (\pm 16)$. The
uncertainties are set to the uncertainty in the observed transit
time, while the values in parenthesis give the uncertainty in the
predicted time. The total uncertainty in the $O-C$ values is the
sum of these two values. This corresponds to a time shifted of
nearly $3~\sigma$ after its expected value. However, it is
difficult to draw conclusion from this result as the determination
of mid-transit times are extremely sensitive to fitting the
ingress and egress of the eclipse profile. Consequently,
correlated noise or a nonuniform stellar brightness distribution
could change the ingress/egress shapes of the transit and thus the
central time away from what is expected for a uniform stellar
disk. A recent study uses seven transits and seven eclipses of this exoplanet to derive a precise set of transit time measurements, with an average accuracy of 3 seconds and reveal a lack of transit-timing variations (Algol et al. 2010).

\subsection{Limb darkening}
\label{sec:ld}

Southworth (2008) shows that the determination of the light curve
parameters, especially the radius of the planet, can be sensitive
to the applied limb darkening model and its coefficients, yielding
a possibility of systematic errors. This is particularly important
for transits observed in the visible wavelength, but less important for transits observed in the infrared. Nevertheless, we investigated the importance of stellar limb darkening on the
light-curve solutions and parameter uncertainties. We compared our
results obtained from fits with various limb darkening laws and
different numbers of fit coefficients.

\begin{figure*}
\centering
\includegraphics[angle=90,width=18cm]{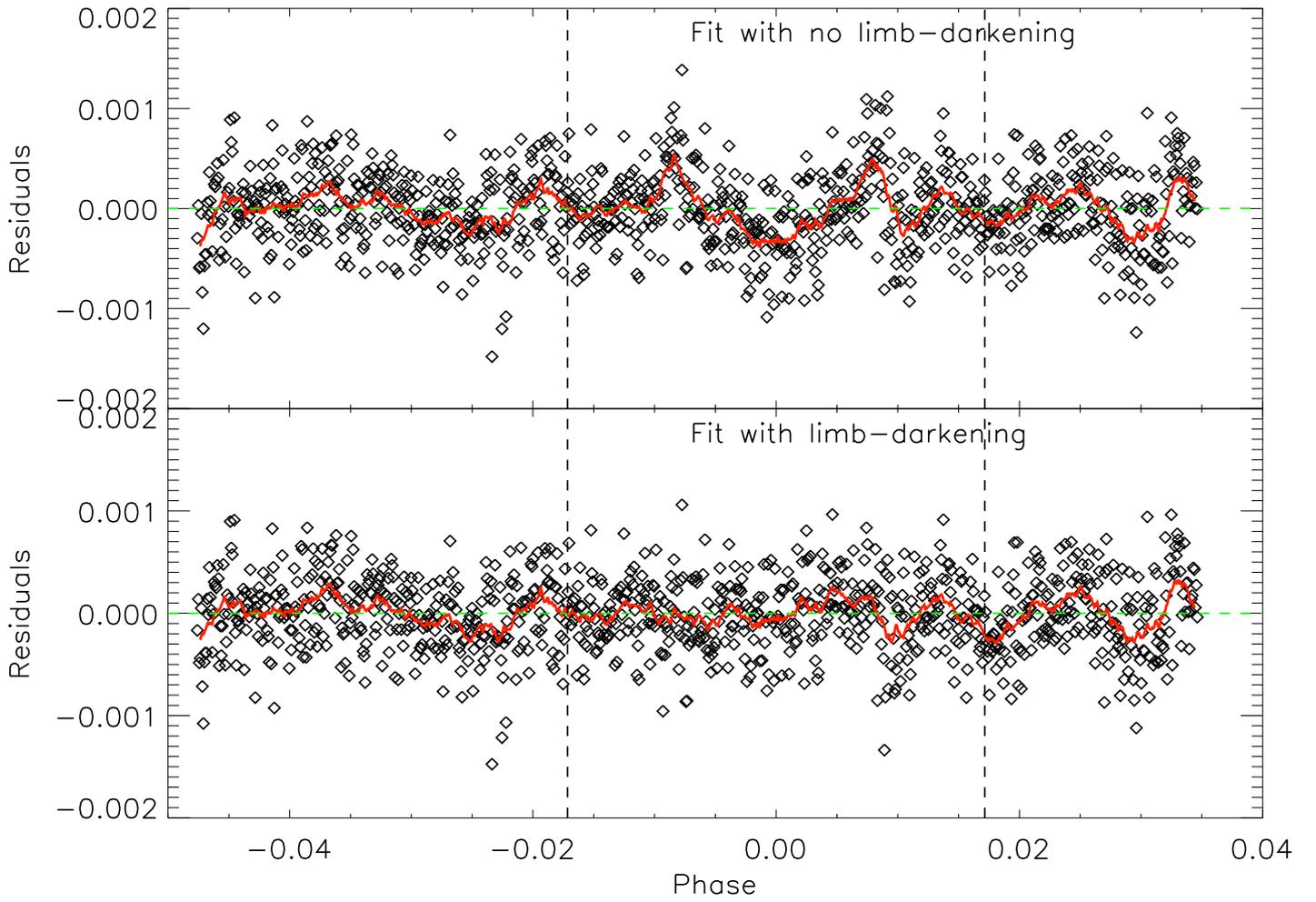}
\caption{Residuals from fits with or without limb darkening
corrections (bottom and top panel respectively). Data (diamonds)
are binned by 120 (~15 seconds) and a smooth of these binned residuals is
overplot in red. Vertical dashed lines mark the beginning and
the end of the transit (first and last contact). The fit is better with limb darkening. No
signature remains in the residuals.} \label{fig:residus}
\end{figure*}

We first note that the fit is clearly improved when including limb
darkening (Fig.~\ref{fig:residus}). We found that accounting for
the effects of limb-darkening  at 3.6~\micron\ decreased the
resulting best-fit transit depth by $(\Delta R_p/R_\star)_{3.6~\mu
m} = 0.0005$ which corresponds to $5~\sigma$ in the present
result. We used a non linear limb darkening law with three fixed
limb darkening coefficients (Table~1) which is slightly different
from the one used in D\'esert et al. (2009). In the present work,
we set $c_1$ to zero and calculate new sets of coefficients c$_2$,
c$_3$ and c$_4$, resulting in a three-parameter non-linear limb
darkening law,
\begin{equation}  \frac{I(\mu)}{I(1)}=1 - c_2(1 - \mu) - c_3(1 - \mu^{3/2}) - c_4(1 - \mu^{2}) \end{equation}
The $c_1(1 -\mu^{1/2})$ term was removed because it reflects the
intensity distribution at small $\mu$ values and is not needed
when the intensity at the limb is desired to vary linearly at
small $\mu$ values (for further details see Sing et al. 2009 and Sing et al. 2010\footnote{http://www.astro.ex.ac.uk/people/sing/}.)
Compared to the quadratic law, the added $\mu^{3/2}$ term provides
the flexibility needed to more accurately reproduce the stellar
model atmospheric intensity distribution at near-infrared and
infrared wavelengths. The limb darkening coefficients for the
three-parameter non-linear law were computed using a Kurucz ATLAS
stellar model\footnote{http://kurucz.harvard.edu/} with
T$_{eff}$=5000 K, log g=4.5, and [Fe/H]=0.0 in conjunction with
the transmission through the IRAC filters and fitted the
calculated intensities between $\mu$=0.05 and $\mu$=1
(Table~\ref{table:ldtab}). This limb darkening law is slightly
different from the one used in the previous analysis (D\'esert et
al. 2009) particularly at the limb ($\mu < 0.05$). Nevertheless,
the signal-to-noise ratio of our data is not high enough to
distinguish the changes between the two sets of limb darkening
coefficients since the difference in the normalized transit light
curves is less than $10^{-6}$.

The photometric precision of our transit light curve allowed us to
perform fits using a non-linear limb-darkening model with the
linear coefficient ($c_2$) as a free parameter. We find
c$_2=1.12994 \pm 0.0243$ in agreement with the value quoted in
Table~\ref{table:ldtab}. As a test, we also fitted the transit
light curve using a linear limb darkening law with one single
coefficient c$_1$ as free parameter. We derived c$_1=0.27821 \pm
0.0291$ and find that $R_p/R_\star$ is $2.5 \sigma$ above the
solution obtained with a the non-linear limb darkening law, but
with a less good $\chi^2$, showing that this solution can be
excluded and the non-linear law is preferred.

\subsection{Comparison with previous observations}

\begin{figure*}
\centering
\includegraphics[angle=90,width=18cm]{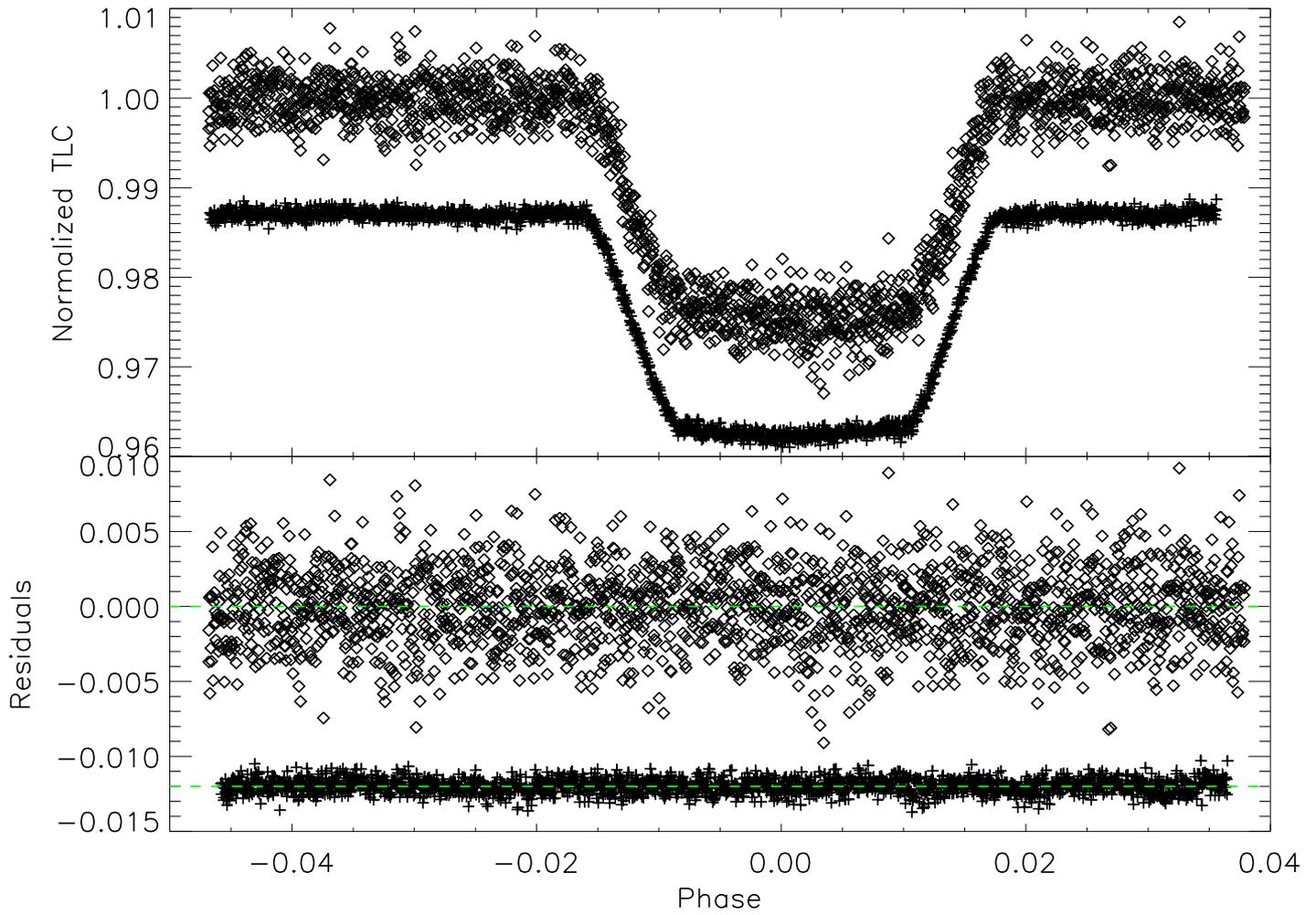}
\newline
\caption{Top panel: normalized transit light curves obtained in
Stellar mode (diamonds) and in Subarray mode (crosses). The
Subarray mode transit light curve is shifted vertically for display purpose. The
Subarray mode transit light curve is binned by 64 to obtained the
same number of data points ($\sim 1900$ points) as the Stellar
mode transit light curve. Bottom panel: residuals obtained from
the fits of the transit light curve. Residuals from the binned
Subarray mode transit light curve are shifted vertically (for
display) and are 4 times smaller.} \label{fig:tlcallmode}
\end{figure*}

We compare the parameters derived here with the results obtained
from visit~1 and 2 observations (D\'esert et al. 2009). Results
from fits to the transit light curve obtained in visit~1
observations obtained at 3.6~\micron\ are summarized in
Table~\ref{table:resultstab}. We measured a dispersion of $2.1
\times 10^{-3}$ (rms) on the individual $\sim 1900$ data points
with an exposure time of 0.4 s. This corresponds to a $\sn$ of
400:1 per image.
To compare with our present data set, we binned the visit~3
transit light curve by $64$ and obtained $\sim 1900$ points with a
$rms = 4.9 \times 10^{-4}$. The two transit light curves, obtained
for visit~1 and 3 at 3.6~\micron, are overplotted with 1900 points
each in Fig.~\ref{fig:tlcallmode}. The $S/N$ by subarray frame is
four times higher than by stellar mode exposure, in agreement with
expected photon noise for a total exposure time sixteen times
longer.

\begin{figure*}
\centering
\includegraphics[angle=90,width=18cm]{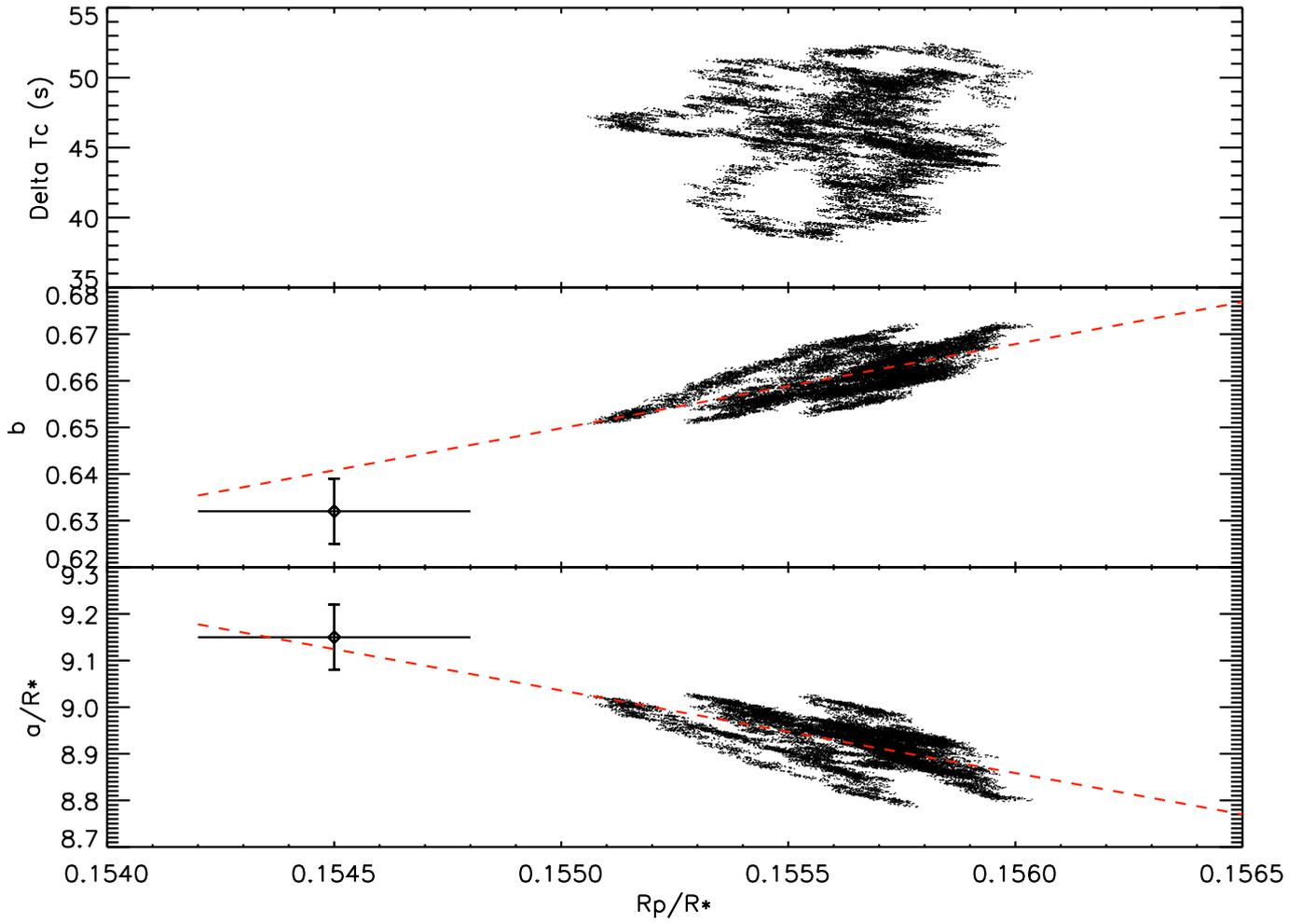}
\caption{Joint distributions of parameters obtain from the Prayer
Bead method using the subarray mode observations. The x-axis
corresponds to radius ratios for all the panels. The top panel is
the fit of the central time, the middle and bottom panel
corresponds to the impact parameter and the scale of the system
respectively. The diamonds with error bars correspond to the
results obtain from visit~1 observations (D\'esert et al. 2009).
The dashed red lines correspond to linear fits of the points from
visit~3, and reveal the correlations between $R_p / R_\star$, $b$
and $a / R_\star$.}
\label{fig:jointdistrib}%
\end{figure*}

\begin{figure*}
\centering
\includegraphics[angle=90,width=18cm]{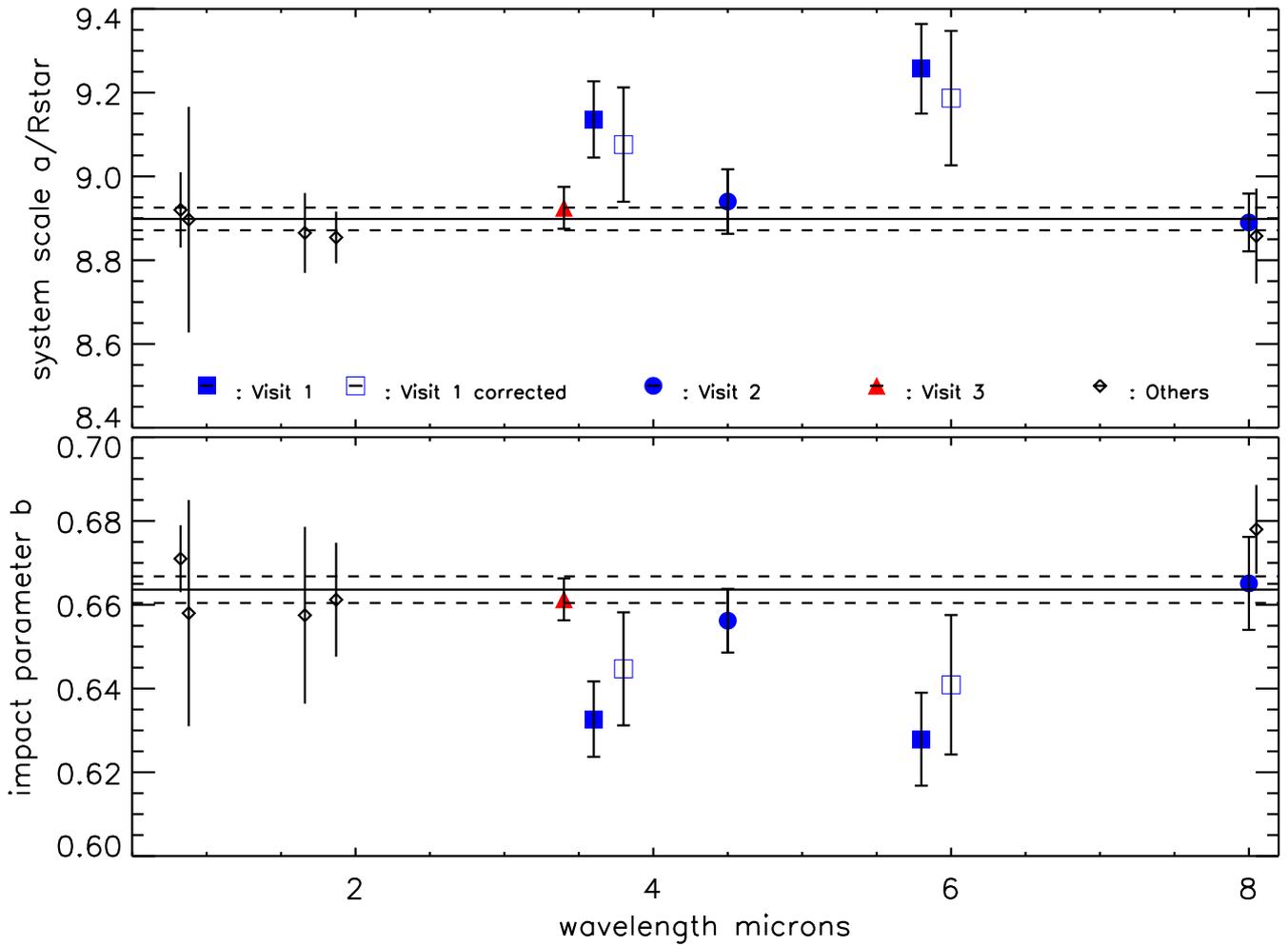}
\newline
\caption{Orbital parameters derived from several studies. Top
panel: impact parameter $b$,  Bottom panel: system scale $a /
R_\star$. Diamonds are values from several other studies: data are
from Pont et al. (2007) in the visible,  from Sing et al (2009) in
the near infrared and from Knutson et al. (2007) at 8.0\micron\ also measured by Carter \& Winn (2010).
Blue squares are from our visit~1 and 2 observations in the four
IRAC channels (D\'esert et al. 2009). The red square are from
visit~3 observations (this work). Continuous and dashed straight
black lines indicate the average and error values obtained from
data excluding outliers. These values are considered as best
estimates for $b$ and $a / R_\star$. Parameters derived from
visit~1 at 3.6 and 5.8~\micron\ respectively, are significantly
different from the best estimates. The empty squares corresponds
to the derived parameters using to the visit~1 observation and
transit light curve model which take into account occulted spots
during ingress or egress.} \label{fig:orbitparams}
\end{figure*}

We are particularly interested in the comparison of the ratios of
the radius of the planet over the radius of the star at
3.6~\micron\ in the two observation sets. The radius ratio found
in visit~3 stands $4\sigma$ above the value derived from visit~1
at 3.6~\micron\ (Table~\ref{table:resultstab}).
The impact parameters $b$ and system scales $a/R_\star$ measured
in the transit light curves provide informations for interpreting
these results.
We found that the $b$ and $a/R_\star$ values derived from visit~1
observations at 3.6~\micron, are respectively $3\sigma$ below and
above the values obtained here at the same wavelength with visit~3
observations (Fig.~\ref{fig:jointdistrib}
and~\ref{fig:orbitparams}). The same results are obtained for
visit~1 observations at 5.8~\micron\ (both observations at
3.6~\micron\ and 5.8~\micron\ were accumulated simultaneously
during visit~1). As already noticed in D\'esert et al. (2009), the
$b$ and $a/R_\star$ measured with visit~1 and 2 stellar mode
observations are consistent between channels observed
simultaneously, but disagree between channels observed at two
different epochs. This suggests that unrecognized systematics
remains between the observations obtained at the two epochs. This
systematic effect could be either instrumental or astrophysical.
Below, we estimate the impact of those systematics on the radius
determination, and possible astrophysical origins of these
discrepancies.

The value of the impact parameter $b$ has a direct effect on the
derived planetary radius. This can be seen theoretically from the
analytic approximation (Carter et al. 2008):

\begin{eqnarray}
    b^{2} = 1- \frac{T}{\tau} \times \frac{R_p}{R_\star}
\label{eq:b}
\end{eqnarray}

where $\tau$ is the ingress or egress duration($\tau = t_{\rm II}
- t_{\rm I}$), $t_{\rm I}$ and $t_{\rm II}$ are the time of first
and second contact, and $T$ is the total transit duration.
Consequently, underestimating the $b$ from $0.66$ to $0.63$,
without changing the timing of ingress and egress would lead to
overestimate $R_p/R_\star$ by $0.001$, which corresponds to
$10~\sigma$ in the result from visit~3.

We estimated the influence of systematic errors on the estimate of
$b$ on the determination of the radius.

We fixed the $b$ value of visit~1 to the one obtained at higher
$S/N$ in visit~3 ($b=0.661$) and fitted visit~1 transit light
curves at 3.6~\micron\ and 5.8~\micron. In that case, for visit~1,
we obtained $\ratioone=0.15532$; these values are in
agreement with the results obtained in visit~3 (Fig.~\ref{fig:orbitparams}).

In conclusion, we find consistent values for $R_p/R_\star$ and
$a/R_\star$ in all data sets if the impact parameters $b$ is set
to the same value. Our final value is in agreement with the values
$b=0.671 \pm 0.008$ and $a/R_\star=8.92 \pm 0.09$ derived by Pont
et al. (2008) with \emph{HST/ACS}, with the values obtained by
Sing et al. (2009) with \emph{HST/NICMOS}, and with $a/R_\star=8.924 \pm 0.022$ from Carter \& Winn (2010) using several Spitzer/IRAC transit light curves at 8~\micron. This suggests that $b$
and $a/R_\star$ derived from visit~1 are affected by unknown systematics which could be either instrumental or astrophysical
such as starspots (see next section~\ref{sec:starspots}).

\subsection{Impact of starspots on the derived parameters}

\label{sec:starspots}

HD 189733 is an active K star, which has been observed to vary
photometrically by $\pm1.5$\%~at visible wavelengths (Henry \& Winn \ 2008; Croll et al.\ 2007; Miller-Ricci et al.\ 2008). Those variations are caused by rotational modulations
in the visibility of star spots on a rotation period of $11.953
\pm 0.009$ days (H\'ebrard \& Lecavelier des Etangs 2006; Henry \&
Winn 2008).

In this section, we consider the possibility that transit light
curves can be modified by the presence of spots on the surface of
the star, introducing systematic errors on the parameters
estimates. In particular, we investigate the possible contribution
of stellar spots and brightness inhomogeneities to explain the
discrepancy between the planet radii at 3.6~\micron\ measured in
visits~1 and 3 data (Sect.4.5.2). We also investigate the effect
of spots to explain the bias in the orbital parameters estimated
using visit~1 data (Sect.~\ref{sec:spotcorr}).

\subsubsection{Ground based photometric follow-up}
\label{sec:ground_based}

We use ground based photometric follow-up to estimate and to
correct for the stellar activity at the epochs of our 3 visits
observations.

\begin{figure*}
\centering
\includegraphics[angle=90,width=12cm]{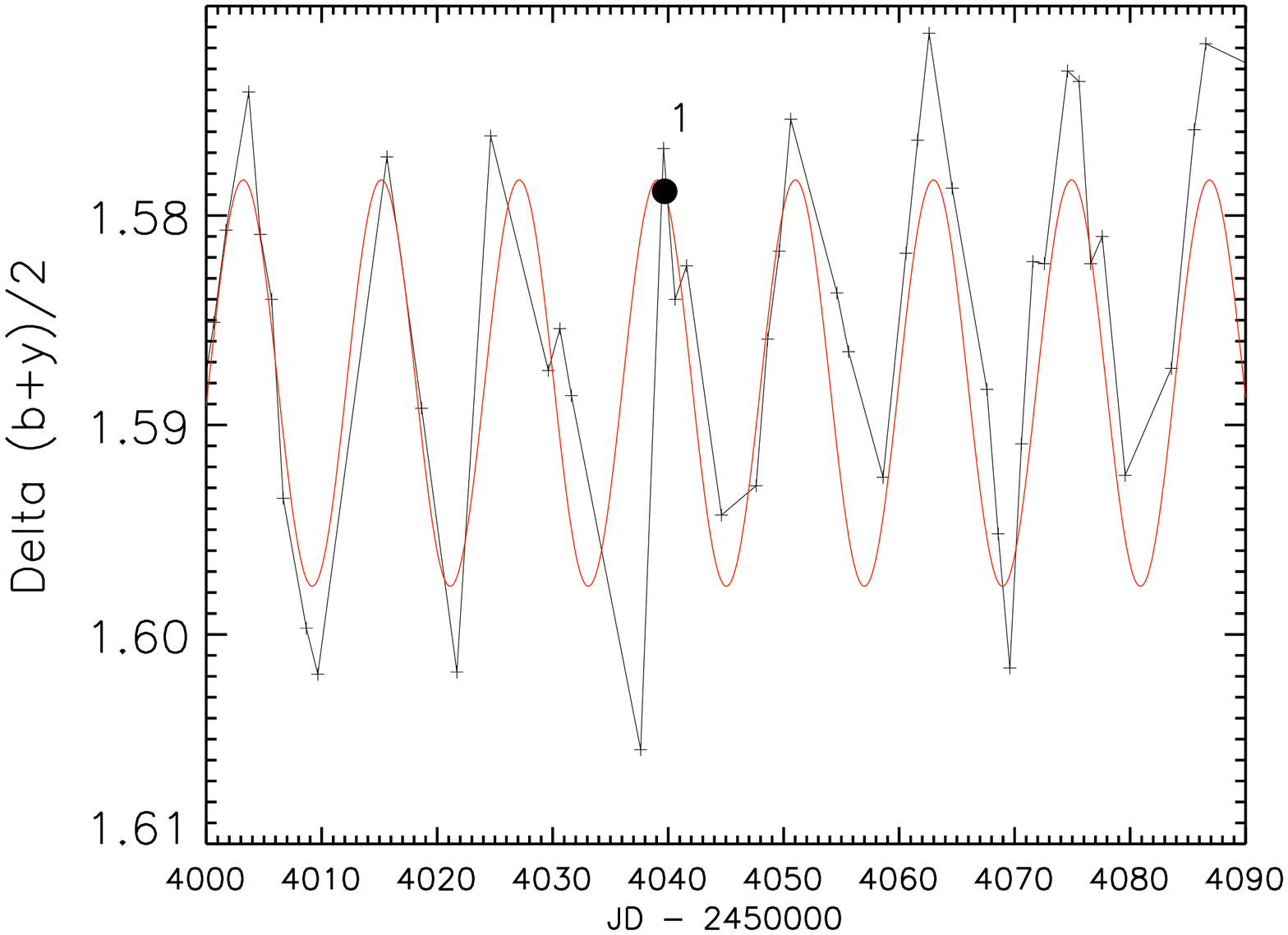}
\includegraphics[angle=90,width=12cm]{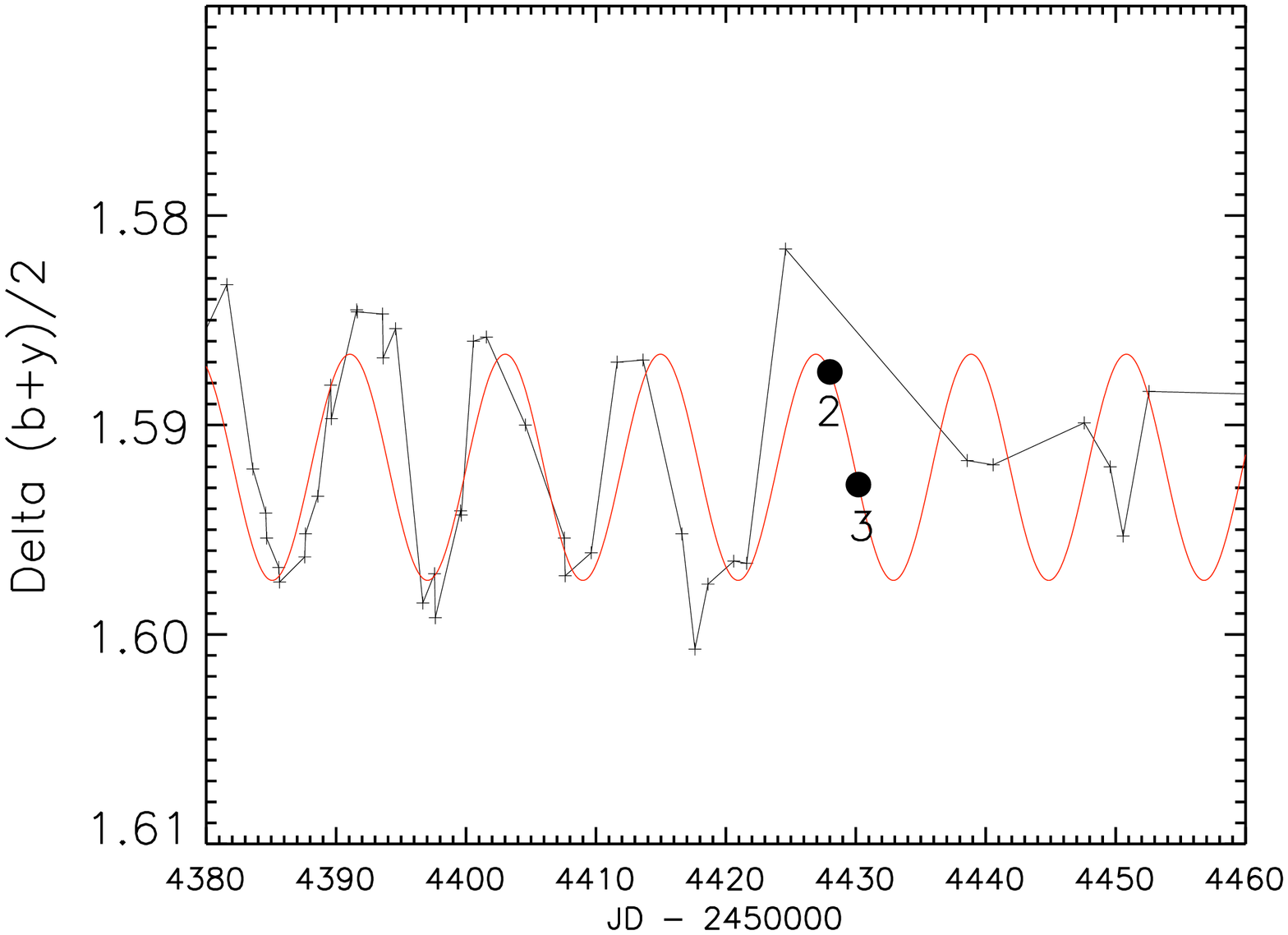}
\caption[]{ {\bf a. (top)}. Relative flux variation for HD 189733
observed with ground-based follow-up encompassing the time of our
IRAC observations (connected crosses). Black circular spot
indicates the time of our \spitzer\ Stellar mode observations at
3.6 and 5.8~\micron\ (Visit~1). Variations are caused by rotational
modulation in the visibility of star spots with a rotation period
of 11.953 days (Henry \& Winn 2008). The continuous curve
overplotted in red corresponds to a sinusoid with a period of
11.95 days. {\bf b. (bottom)}. The same as above for the stellar mode observations at 4.5 and 8.0~\micron\ (Visit~2) and Subarray mode observations at 3.6~\micron\ (Visit~3).}
\label{fig:groundbased}
\end{figure*}

The star has been followed-up in V-band over several years using
the Tennessee State University $0.8~m$ automated photometric
telescopes (APT) at Fairborn Observatory (Henry et al. 2008).
These data show that the stellar flux variation reach a peak to
peak maximum of $3\%$ in the visible (Fig.~\ref{fig:groundbased}).

The observed variations in the visible have to be translated to
corresponding variation in the infrared. Let's define
$f_{\lambda}$, the relative variation in flux due to stellar
activity:
\begin{eqnarray}
f_{\lambda} \equiv \frac{F_\star(\lambda) - F_{quiet}(\lambda)
}{F_{quiet}(\lambda)}, \label{eq:s}
\end{eqnarray}
\noindent where $F_\star$ is the measured flux and $F_{quiet}$ is
the reference stellar flux (e.g., without stellar activity).
$f_{\lambda}$ is negative when the star is fainter and positive
when it is brighter than the reference brightness.

Previous HST ACS observations of HD 189733 have established that
its stellar spots have an effective temperature approximately
1000~K cooler than that of the stellar photosphere (Pont et al.
2008) with spot size $\sim 2.8\%$ of the stellar disk area.
Following Knutson et al.\ (2009), we considered that the amplitude
of the stellar flux variations scales approximately as the ratio
of the flux of blackbodies at 5000 and 4000~K. Thus, we used the
normalized difference of Planck functions to estimate the maximum
variation of flux at 3.6\micron\ $f_{(3.6\mu m)}$. The $3\%$
maximum variation in the V-band due to the presence of spots on
the stellar surface, scales to $f_{(3.6\mu m)} \sim 0.8\%$ in the
infrared.

The ground-based photometry suggests that visit~1 happened during
the maximum brightness of the star (Fig.~\ref{fig:groundbased}).
Unfortunately, no ground based observations were acquired during
visit~2 and 3. Thus, we have no direct measurement of the stellar
activity at those precise epochs. Nevertheless, the activity
period $P=11.953\pm0.009$ (Henry \& Winn 2008) can be used to
interpolate the star brightness at the epochs of visit~2 and 3. We
found that the star should be close to its maximal brightness
during visit~2, and $\sim 2\%$ below the maximal brightness during
visit~3, corresponding to $f_{(3.6\mu m)} \sim -0.5\%$ at this
epoch.

\subsubsection{Stellar spots outside the zone occulted by the planet}
\label{sec:unoccult}

Spots and bright faculae cause variations in the star brightness
($F_\star$). Nonetheless, if the surface brightness of the zone
occulted by the planet is not modified, the absolute depth of the
transit light curve ($\Delta F_\star$) remains the same.
Therefore, at epoch of low stellar brightness like during visit~3,
the relative transit depth ($\Delta F_\star/F_\star$) can be
significantly overestimated, leading to an overestimate of the
planet to star radius ratio.

As seen in Corot-2b (Czesla et al.\ 2009), the transit depth
constraining the planet radius ratio is correlated with the star
brightness. We can empirically define $\alpha$ by:
\begin{eqnarray}
\frac{\left(\frac{R_p}{R_\star}\right)_{Measured}^2 -
\left(\frac{R_p}{R_\star}\right)_{True}^2}{\left(\frac{R_p}{R_\star}\right)_{True}^2}
= \alpha f_{\lambda}, \label{eq:alpha}
\end{eqnarray}
\noindent where $(R_p/R_\star)_{True}$ is the true radius ratio
which could be obtained from the transit light curve when the star
presents no spots and is at the reference brightness. The relative
error on the measured radius ratio is therefore given by :
\begin{eqnarray}
\frac{\Delta(R_p / R_\star)}{\left(R_p / R_\star\right)_{True}}
\approx \frac{\alpha f_{\lambda}}{2}, \label{eq:deltarp}
\end{eqnarray}

\begin{figure*}
\centering
\includegraphics[angle=90,width=20cm]{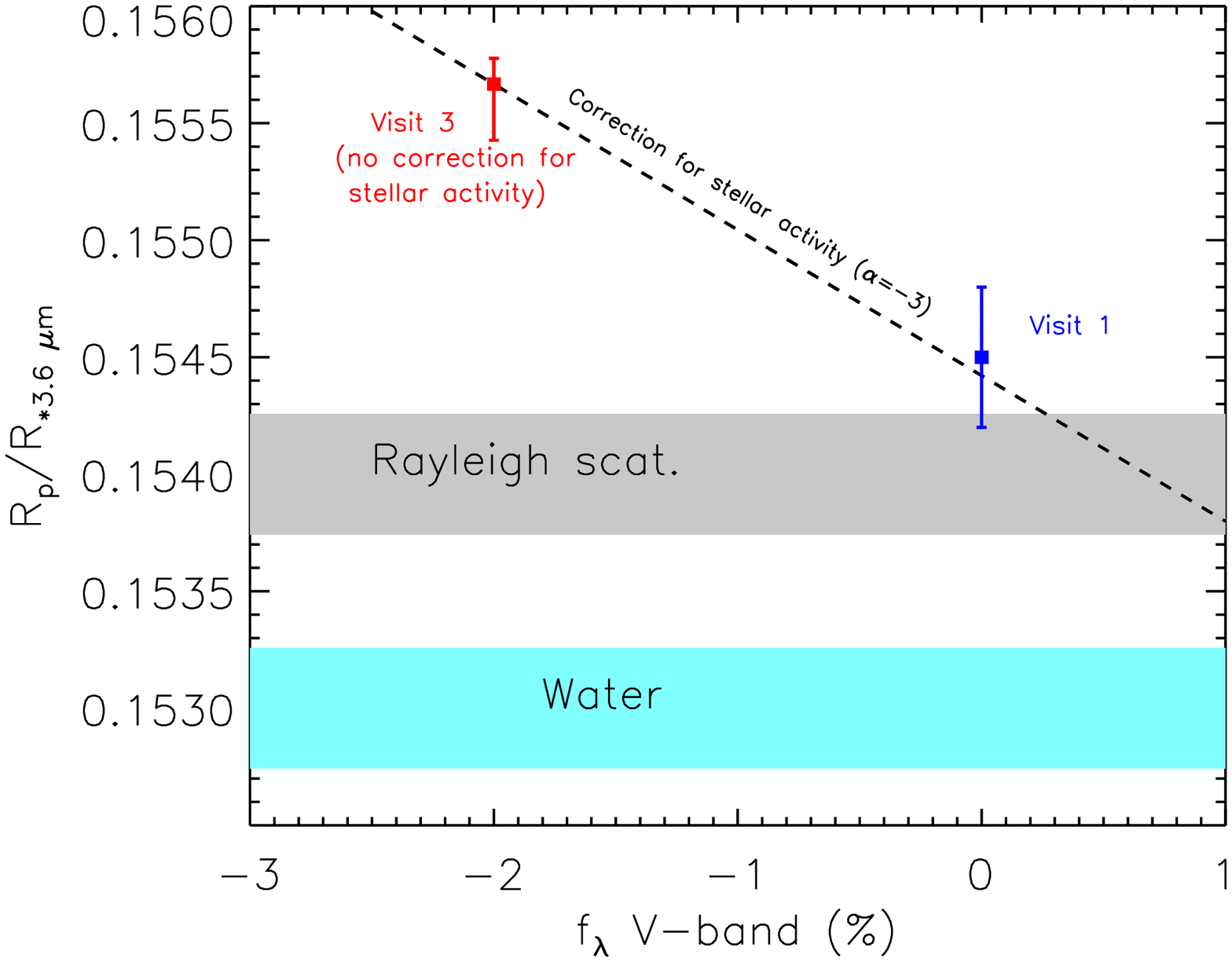}
\caption{Measured values of $R_p/R_\star$ ratios at 3.6~\micron\
as fonction of the relative decrease of stellar flux due to spots
(in V-band). The blue and red squares are the $R_p/R_\star$
derived from visit~1 and 3 observations respectively. The dotted
line corresponds to the theoretical correction of the relative
difference between the measured radius ratio and the expected
value as fonction of the relative decrease of stellar flux in the
V-band $f_{\lambda}$. Using ground based photometric follow-up
(Fig.~\ref{fig:groundbased}), we set $f_{(3.6~\mu m)}$ to zero for
visit~1 (see text~\ref{sec:ground_based}). The $R_p/R_\star$ we
derive from visit~3 observations (no correction) is above the one
obtained in visit~1 ($4~\sigma$). To compare both visits, the
$R_p/R_\star$ derived from visit~3 has to be corrected for spots
(Sect.~\ref{sec:spotcorr}). Simultaneous observation in V-band
shows that $f_{V-band}=-2\%$ during visit~3
(Fig.~\ref{fig:groundbased}). $(R_p/R_\star)_{Corrected}$ for
visit~3 is in agreement with $R_p/R_\star$ derived in visit~1 for
a slope $\alpha=-3$.} \label{fig:fig_comparaison_mode}
\end{figure*}

In Corot-2b data (Czesla et al.\ 2009), we found $\alpha\sim 1.7$.
With the hypothesis that the stellar surface brightness outside
the spot areas is not modified by the stellar activity, we
obtained $\alpha=-1$. With the photometric variations obtained
from ground based observation at epochs close to the Spitzer
observations, we found $f_{3.6\mu m}=-0.5\%$ in visit~3 compared
to visit~1. Therefore, the radius ratio $R_p/R_\star$ derived from
visit~3 which is $\sim 0.75\pm 0.18\%$ above the radius derived
from visit~1 (Table~\ref{table:resultstab}) can be explained using
equation~\ref{eq:deltarp} if $\alpha \sim -3\pm 1 $ (Fig.~\ref{fig:fig_comparaison_mode}).

In short, the discrepancy between the visit 1 and 3 radius ratios
at 3.6~\micron\ can be explained by the effect of stellar activity
and the presence of spots during the visit~3.

\subsubsection{Occulted starspots}
\label{sec:spotcorr}

The planet occulting dark or bright areas on the surface of the
star during transit produces a rise or decrease in the flux.
Because the ingress and egress parts of the transit light curve
strongly constrain the orbital parameters of the planet, dark or bright areas occulted by the planet during the ingress or egress change the
apparent time of the transit contacts, introducing systematic
errors in the derived $b$ and $a/R_\star$ parameters
(Eq.~\ref{eq:b}). Consequently, the low $b$ value obtained in
channel~1 and~3 using visit~1 data could be caused by the presence
of cold spots or bright faculae occulted by the planet.

To estimate the influence of occulted star cold or bright regions
during the ingress and egress on the determination of our
parameters, we fitted the visit~1 and 2 observations with a
transit light curve model including occulted star spots (cold or
hot) during ingress and egress. These fits have a goodness of fit
equivalent to the fits using the model without spots. For visit~2,
the model with occulted spots give similar results as D\'esert et
al. (2009). Most importantly, for visit~1, the model fits well the
data with a bright spot occulted during the ingress part of the
light curve. The best fit is found assuming a spot size similar to
the size of the planet and a spot surface brightness $8\%$
brighter than the star. Compared to the results given in D\'esert
et al.\ (2009), this new model applied to visit~1 data at
3.6~\micron\ gives a larger $b$ by 0.013 ($2~\sigma$), a smaller
$a/R_*$ by 0.06 ($1~\sigma$), and error bars larger by a factor of
about $1.5$. Interestingly, the $R_p/R_*$ found with this new
model changes by less than $0.3~\sigma$. Similar conclusion are
obtained with data at 5.8~\micron\ (Fig.~\ref{fig:orbitparams}).

As a conclusion, the surprising results for $b$ and $a/R_*$ found
by D\'esert et al.\ (2009) in visit~1 data can simply be explained
by astrophysical systematics in the transit light curve due to
stellar activity. However, the difference in the measured values for the radius ratios $R_p/R_\star$ between Visit~1 and 3 cannot be explained by such dark and bright areas occulted during ingress or egress. It must be caused by the fluctuation of the whole stellar disk brightness (see Sect.~\ref{sec:unoccult}).

\subsection{Opacities in the IRAC bandpasses }

\begin{figure*}
\centering
\includegraphics[angle=90,width=20cm]{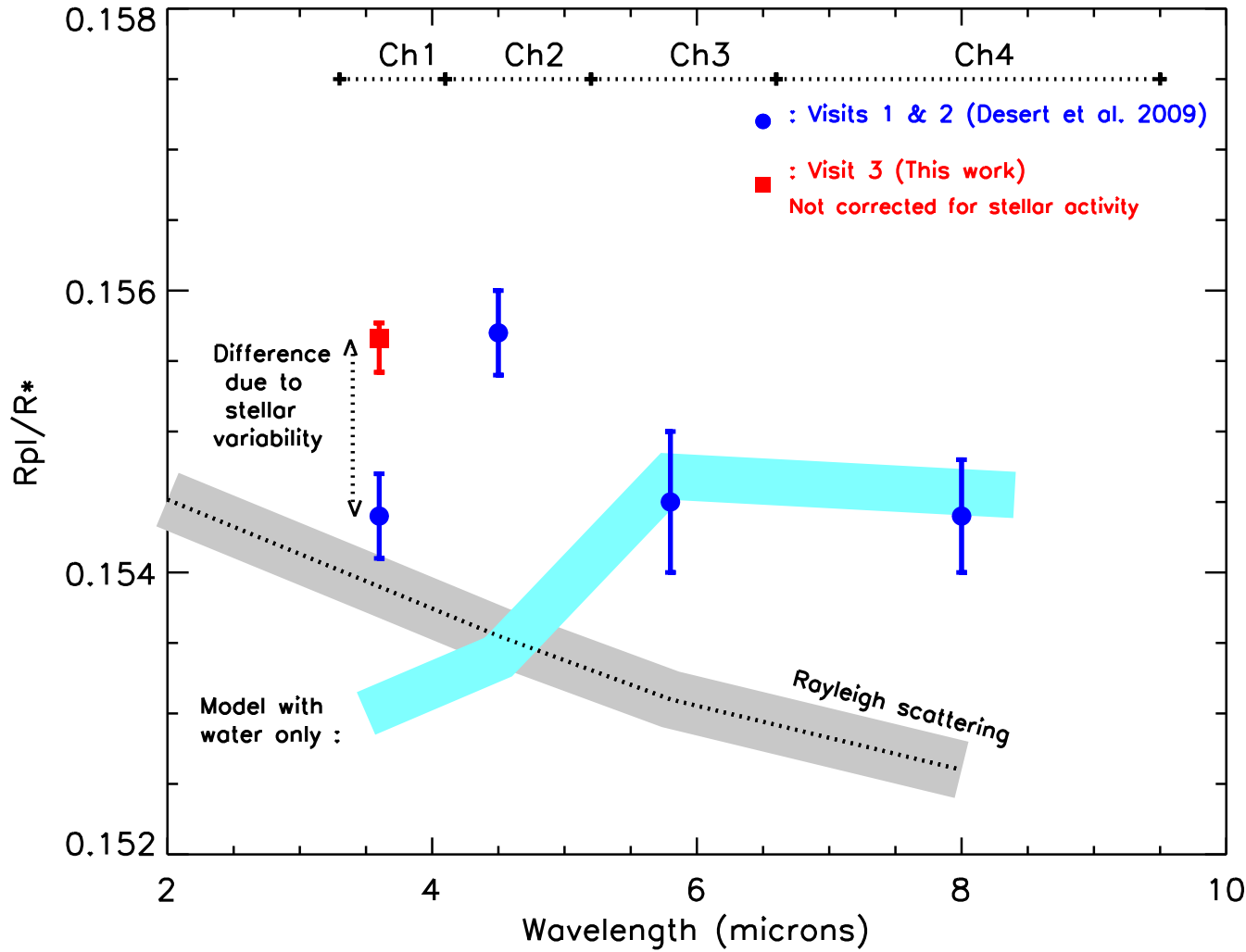}
\caption{Census of the measured values of $R_p/R_\star$ ratios in
the four \spitzer/IRAC channels for \hd\ obtained using three transit observations during visit~1 and 2 (blue circles; Désert et al. 2009), and visit~3 (red square; this work) with their $1~\sigma$ error bars. The data point obtained for visit~3 has to be corrected from stellar spots (see
text~\ref{sec:spotcorr}). The IRAC bandpasses for each channel are shown at the top with dotted line. The theoretical planet-to-star radius ratios with absorption by only water molecules or by only  Rayleigh scattering are plotted in light blue and grey respectively.} \label{fig:radii}
\end{figure*}

The $R_p/R_\star$ measured at 3.6~\micron\ from visit~1 is $4~\sigma$ above the value measured in visit~3. This difference can be reconciled by correcting for the $2~\%$ stellar variability monitored with the ground-based observations between the two visits.
However, at least two colors measured simultaneously are required to ensure that the measured variations of the wavelength dependant radius of the planet are caused by the presence of atmospheric chemical absorbers.
In this framework, the $R_p/R_\star$ measured at 3.6~\micron\ from visit~1 is $4~\sigma$ above the opacity due to water if we consider that the radius simultaneously measured at 5.8~\micron\ is due to water opacity.
Therefore, other species beyond \HOO\ must be present in the
atmosphere of the planet and absorb at 3.6~\micron. 
One possible physical explanation for the large radius at 3.6~\micron\ could be due to Rayleigh scattering (Lecavelier des Etangs et al. 2008b). Effectively, the radius mesured at this wavelength is in agreement at one sigma level with its expected value if we consider Rayleigh scattering opacity (Fig.~\ref{fig:radii}). We note that the possible presence of a strong methane (\CHQ) band at 3.3~\micron\
could also contribute to the opacity in this bandpass (Sharp \&
Burrows 2007). The radius we derived at 4.5~\micron\ could be
interpreted as due to CO molecules which exhibit a large opacities
in this bandpass (D\'esert et al. 2009). The radii we derived at
5.8~\micron\ and 8.0~\micron\ are at more than $3~\sigma$ above
the absorption due to Rayleigh scattering, and could be
interpreted by the presence of water molecules which opacity
overcomes Rayleigh scattering at those wavelengths
(Fig.~\ref{fig:radii}). These two bandpasses exhibit approximatly
the same radius value, consistent with a model which includes
water. 

Therefore water vapour could be present in the atmosphere of this hot-Jupiter ; but it cannot be evidenced  simply comparing planetary radii measured at 3.6 and 5.8~\micron, because the radius measured at  3.6~\micron\ is affected by some other species absorption.

\section{Conclusion}

We obtained a new high-$S/N$ photometric transit light curve
at 3.6~\micron\ using the \emph{IRAC/Spitzer} subarray mode
observations. This observing mode offers a higher photometric cadence, therefore a better signal-to-noise ratio than the previous measurements (stellar mode) obtained at this wavelength. We derive a value of $R_p/R_\star=0.15566 \pm
0.00011$ at 3.6~\micron\ which is $4\sigma$ above the value
derived from previous observations in the same bandpass (D\'esert
et al.\ 2009). This difference is explained by the stellar variability
between the two observational epochs. Simultaneous ground based
observation allowed us to correct for these variations. A photometric variation of $\sim 2\%$ in V-band was monitored between the two measurements and provide a correction of $-0.75\%$ for the new 3.6~\micron\ measurement. Noteworthy, the first observation was obtained during the maximum brightness of the star, which led to the minimal possible value of the measured $R_p/R_\star = 0.1545 \pm 0.0003$ at 3.6~\micron\ for this planet (D\'esert et al. 2009). This final planet-to-star radius ratio at 3.6~\micron\ is above the expected value if only water molecules were contributing to the absorption at both 3.6 and 5.8\micron.

We showed the importance of long term ground based monitoring program in order to take into account the stellar
activity and to compare results obtained from observations carried out
at different epochs. Any estimate of the planet to star radius
ratio should be associated with a corresponding stellar activity
level.

\newpage


    \begin{table}
    \begin{tabular}{c c c c c}        
    \hline\hline                 
    Visit & C1 & C2 & C3 & C4  \\
    \hline\hline                        
     \\
      1 (D\'esert et al. 2009)   & 0.6023  & -0.5110 & 0.4655   & -0.1752  \\
      3 (This work) & 0.0000 & 1.13694 & -1.3808 & 0.55045  \\
           \hline                                   
     \end{tabular}

\caption{Limb darkening coefficients used for IRAC/channel~1
bandpass centered at 3.6~\micron. The coefficients are the ones
used for visit~3 (this study) and used in our previous analysis
(visit~1; see D\'esert et al. 2009). The two sets of coefficients
only differ at the limb (see Sect.~\ref{sec:ld})}
    \label{table:ldtab}

     \end{table}

\newpage

\begin{table*}
    \begin{tabular}{c c c c c}        
    \hline\hline                 
      Parameter at $3.6~\mu m$ & Median (visit~3) & $1\sigma (68\%)$ & $2\sigma (95\%)$ & visit~1 (D\'esert et al. 2009)\\    
    \hline\hline                 

     \\
$R_p / R_\star$  &  $0.15566$            & $^{+0.00011}_{-0.00024}$      & $^{+0.00022}_{-0.00043}$ & $0.1545 \pm 0.0003$\\
     \\

    $b$              &  $0.6613$              & $^{+0.0053}_{-0.0050}$      & $^{+0.0078}_{-0.0079}$ & $0.632 \pm 0.007$\\
     \\

    $a / R_\star$    & $8.925$              & $^{+0.049}_{-0.052}$      & $^{+0.079}_{-0.096}$ & $9.15 \pm 0.07$ \\
     \\

    $T_c (s)$        & $47$              & $^{+3.4}_{-3.6}$      & $^{+5.0}_{-6.4}$ & $19 \pm 12$ \\
    \\

    \hline                                   
    \end{tabular}

    \caption{Fitted parameters of the transit light curves at 3.6~\micron\ for Visit~1 and 3. Quoted errors corresponds to 68\% and 95\% of the realizations (Visit~3).}
    \label{table:resultstab}

\end{table*}


\newpage

\newpage

%

\begin{acknowledgements}

D.K.S.\ is supported by CNES. This work is based on observations
made with the \emph{Spitzer Space Telescope}, which is operated by
the Jet Propulsion Laboratory, California Institute of Technology
under a contract with NASA. D.E. is supported by the Centre National d'\'Etudes Spatiales (CNES). 
D.E. acknowledges financial support from the French \emph{Agence Nationale pour la Recherche} through
ANR project NT05-4\_44463.
We thank our anonymous referee and our editor for their comments that strengthen the presentation of our results.

\end{acknowledgements}

\end{document}